\documentclass[aps,pra,preprint,showpacs]{revtex4-1}

\usepackage{amsfonts}
\usepackage{amsmath}
\usepackage{amssymb}
\usepackage{graphicx}

\begin{document}

\title{Electrically tunable magnetoplasmons\\ in a monolayer of silicene or
germanene}
\author{M. Tahir$^{\dag}$ and P. Vasilopoulos}

\affiliation{Department of Physics, Concordia University, Montreal, Quebec, Canada H3G 1M8}

\begin{abstract}
We  theoretically study  electrically tunable magnetoplasmons
in a monolayer of silicene or germanene. We derive the dynamical response
function and take into account the effects of strong spin-orbit coupling (SOC) and of an
external electric filed $E_z$ perpendicular  to the plane of the buckled silicene/germanene.
Employing the random-phase approximation  we analyze the magnetoplasmon
spectrum. The dispersion relation has the same form as in a two-dimensional electron gas with the cyclotron and plasma frequencies modified due to
the SOC and the field $E_z$. In the absence of SOC and $E_z$, our results agree well with recent experiments on  graphene. The
predicted effects could be tested  by experiments similar to those on
graphene and would be useful for future spintronics and optoelectronic
devices.
\end{abstract}

\maketitle

\preprint{ }

\affiliation{Department of Physics, Concordia University, Montreal, Quebec,
Canada H3G 1M8}

\section{INTRODUCTION}

Since its realization as a truly two-dimensional (2D) material,
graphene has attracted much interest, both due to fundamental science and
technological importance in various fields \cite{1,2}. However, the realization of a
tunable band gap, suitable for device fabrications, is still challenging and
SOC is very
weak in graphene. To overcome these limitations researchers have been increasingly studying
similar materials. One such material, called silicene,  is a monolayer honeycomb structure of silicon and has been predicted to be stable \cite{3}.
Already several attempts have been made to synthesize it \cite{4}. A similar material is germanene.

Despite controversy over whether silicene has been experimentally created or not  \cite{5}, it is expected to be  
an excellent candidate materials because it has a strong SOC and an electrically tunable
band gap \cite{6,7,8}. It's  a single layer of silicon atoms with
a honeycomb lattice structure and compatible with silicon-based  electronics that dominates the
semiconductor industry. Silicene has Dirac cones similar to those of graphene and density functional
calculations showed that the SOC gap induced in it is about 1.55 meV 
\cite{6,7}.  Moreover,  very recent theoretical studies predict  the  stability of  silicene
on non metallic surfaces such as graphene  \cite{9}, boron nitride or SiC \cite{10}, and in graphene-silicene-graphene structures \cite{11}. 
Besides the strong SOC, another salient feature of silicene
is its buckled  structure with the A and B sublattice planes separated by a vertical distance $2\ell$ so that inversion symmetry can be broken by 
an external electric field resulting in a staggered potential 
\cite{8}. Accordingly, the energy gap in it
and in germanene can be controlled electrically. Due to
this unusual band structure, silicene and germanene are expected to show 
 exotic properties such as quantum spin- and valley-Hall
effects \cite{8,12,13}, magneto-optical and electrical transport \cite%
{14,15}, etc..

Plasmons are quantized charge excitations due to the Coulomb interaction and a very
important aspect in condensed matter physics not only from a fundamental point
of view but also from a  technological one \cite{16,17,18,19,20}.  
In the presence of a magnetic field they are called magnetoplasmons and have been extensively studied theoretically \cite%
{21,22,23,24,25} and observed experimentally \cite{26,27,28} in graphene.
The study of graphene (magneto)plasmons involves  
spatial confinement of light and enables them to  
operate at terahertz frequencies thus making it a promising material for optoelectronics. Next
to graphene, which has a very weak SOC and no gap if not grown on a substrate, is silicene or germanene with strong SOC and a tunable band gap. So far 
plasmons in them have been studied only in the absence of  a magnetic field \cite{29,30}, \cite{ben}.

The purpose of this work is to study magnetoplasmons in silicene or germanene. We evaluate
the dynamical nonlocal dielectric response function to obtain the
magnetoplasmon spectrum within the random-phase approximation (RPA).
 In particular, we take into account the effect of strong
SOC and of an external electric field $E_z$ applied perpendicular to its plane.
Experiments can be done by incorporating the effects of SOC and
 $E_z$  similar to the recent ones \cite{26,27,28} on gapless graphene. 
In Sec. II we present the basic formalism, in Sec. III the density-density correlation function, and  in Sec. IV the magnetoplasmons. Results and their discussion follow in Sec. V and a summary in Sec. VI.

\section{Model Formulation}

We consider silicene or germanene in the $(x,y)$ plane in the presence of intrinsic
SOC and of an external electric field  $E_z$ applied along $z$ axis in
addition to a magnetic field ${\bf B}=B\hat{z}$. Electrons in silicene obey the 2D Dirac-like Hamiltonian \cite{7,8}
\begin{equation}
H^{\eta ,s}=v_F(\eta \sigma _{x}\Pi _{x}+\sigma _{y}\Pi _{y})+\eta
s\lambda \sigma _{z}+V\sigma _{z}.  \label{1}
\end{equation}
Here $\eta = 1(-1) $ represents the $K$ ( $K^{\prime }$) valley,  $%
V=2lE_{z}$ is the potential due to the uniform electric field $E_{z}$, $2l= 0.046$ nm is the distance between the two sublattice planes,
 and   $\lambda $ = 4 meV the SOC.  For germanene we have $2l = 0.066$ nm and $\lambda = 43$ meV.  Also,  $\sigma _{x}$, $\sigma _{y}$, $\sigma _{z}$) are the Pauli matrices that
describe the sublattice pseudospin, $v_F$   the electron Fermi velocity,    
and  $s=+1 (-1)$ the   up (down) electron spin.
Further, $\mathbf{\Pi }=\mathbf{p}-e\mathbf{A}$ is the canonical momentum 
and $\mathbf{A}$  the vector potential that yields ${\bf B}=B\hat{z}$; we use the Landau gauge $\mathbf{A}=(0, Bx, 0)$. 
After diagonalizing the Hamiltonian we obtain the eigenvalues
\begin{equation}
E_{n}^{\eta ,s}=\pm \big[\hslash ^{2}\omega _{c}^{2}n+V_\xi^2  
\big]^{1/2}%
,\,\,E_{0}^{\eta ,s}=-\eta V_\xi   
\label{2}
\end{equation}
where $V_\xi=V+\xi\lambda$ and $\xi=\eta s$. The corresponding eigenfunctions are
\begin{equation}
\Psi _{n}^{\eta ,s}=\frac{e^{ik_{y}y}}{\sqrt{L_{y}}}\left( 
\begin{array}{c}
-iC_{n}^{\eta ,s}\phi _{n-1} (\bar{x})\\ 
D_{n}^{\eta ,s}\phi _{n}(\bar{x})%
\end{array}%
\right) ,\Psi _{0}^{+,s}=\frac{e^{ik_{y}y}}{\sqrt{L_{y}}}\left( 
\begin{array}{c}
0 \\ 
\phi _{0}(\bar{x})%
\end{array}%
\right) ,\Psi _{0}^{-,s}=\frac{e^{ik_{y}y}}{\sqrt{L_{y}}}\left( 
\begin{array}{c}
\phi _{0} (\bar{x})\\ 
0%
\end{array}%
\right) . \label{3}
\end{equation}
Here  $\omega _{c}=v_F\sqrt{2eB/\hslash }$, $\bar{x}=x-x_0,\,x_{0}=l^{2}k_{y}$, $l=\sqrt{%
\hslash /eB}$ is the magnetic length, and $L_{y}$   
the length of the  silicene or germannene monolayer along the $y$ direction. Moreover,  
$\phi_{n}(x)=e^{-x^{2}/2}H_{n}(x)/\sqrt{2^{n}n!\sqrt{\pi }l}$  and $H_{n}(x)$ are the
Hermite polynomials. $C_{n}^{\eta ,s}$ and $D_{n}^{\eta s}$ are the 
normalization constants  
\begin{equation}
C_{n}^{\eta ,s}= 
[(1\pm V_\xi /E_{n}^{\eta,s})/2]^{1/2},\,\,D_{n}^{\eta s}= 
[(1\mp V_\xi/ E_{n}^{\eta ,s})/2]^{1/2}.  \label{4}
\end{equation}
The energy spectrum given in Eq.
(2) is degenerate with respect to the wave vector $k_{y}$. 
  The eigenfunctions for the $K^{\prime }$ valley can be
obtained from Eq. (3), by interchanging $\phi _{n}$ and $\phi _{n-1}$,  and the
corresponding eigenvalues from Eq. (2) with $\eta=-1$.

\section{Density-density correlation function}

 i) {\it Finite frequencies.} The dynamic and static response properties of an electron system are  
embodied in the structure of the density-density correlation function which we evaluate in the RPA. The
RPA treatment presented here is by its nature a high-density approximation
that has been successfully employed in the study of collective excitations
in 2D graphene-like systems both with and without an applied magnetic field 
\cite{16}- \cite{24}. It has been found that the RPA predictions of
plasmon spectra are in excellent agreement with experimental results 
\cite{26}- \cite{28}. Following this  technique, one can express the dielectric
function as
\begin{equation}
\epsilon (q,\omega )=1-v_{c}(q)\Pi _{0}(q,\omega ),  \label{5}
\end{equation}
where $v_{c}(q)=2\pi e^{2}/\kappa q$ is the 2D Fourier
transform of the Coulomb potential with wave vector $q$ and $\kappa $ 
the effective background dielectric constant. The non-interacting
density-density correlation function is obtained as
\begin{align}
\Pi _{0}(q,\omega )& =\frac{1}{A}\underset{n,n^{\prime
},k_{y},k_{y}^{\prime }}{\sum }[f(E_{n}^{\eta ,s})-f(E_{n^{\prime }}^{\eta
,s})]\left\vert \left\langle \alpha ^{\prime }\mid e^{-i\mathbf{q.r}}\mid
\alpha \right\rangle \right\vert ^{2}  \label{6} \\
& \times \lbrack E_{n}^{\eta ,s}-E_{n^{\prime }}^{\eta ,s}+\hbar \omega
+i\gamma ]^{-1},  \notag
\end{align}
where $A$ is the area of the system and $\left\vert \alpha \right\rangle
=\left\vert n,\eta ,s,k_{y}\right\rangle $. Here $\gamma $ is the the width of the energy levels
due to scattering and is an infinitesimally small quantity in 
samples with high mobility \cite{28}. The matrix element in Eq. (6) is 
 evaluated in the Appendix; the result is
%
%
%
\begin{equation}
\left\vert \left\langle \alpha ^{\prime }\mid e^{-i\mathbf{q.r}}\mid \alpha
\right\rangle \right\vert ^{2}=J_{n,n^{\prime }}(u )=\delta _{k_{y,}^{\prime
}k_{y-}q_{y}}\Big\{[C_{n}^{\eta ,s}C_{n^{\prime }}^{\eta
,s}]F_{n-1,n^{\prime }-1}(u)+[D_{n}^{\eta ,s}D_{n^{\prime }}^{\eta
,s}]F_{n,n^{\prime }}(u)\Big\} ^{2},  \label{7}
\end{equation}
where $u =l^{2}q^{2}/2$. For $n\leq n^{\prime }$ we have $[F_{nn^{\prime }}(u )]^{2}=(n!/n^{\prime }!) e^{-u }u
^{n^{\prime }-n}[L_{n}^{n^{\prime }-n}(u )]^{2}$ and  for $n^{\prime }\leq n$ 
the same expression with $n$ and $n'$ interchanged.
The sum over $k_{y}$ in Eq. (6) can be evaluated using the prescription ($k_0=L_{x}/2l^{2}$)
\begin{equation}
\sum_{k_y} \rightarrow \frac{L_{x}}{2\pi }g_{s}g_{v}
\int_{-k_0}^{k_0}dk_{y}=\frac{A}{D_{0}}
g_{s}g_{v},  \label{8}
\end{equation}%
where $D_{0}=2\pi l^{2}$, $ g_{s}$ and $g_{v}$ are the spin and valley degeneracies,
respectively. We use $g_{s}$ = $g_{v}=1$ in the present work due to the lifting
of the spin and valley degeneracies in silicene or germanene.

We now use  the transformation $k_{y}\rightarrow-k_{y}$ and the fact  
that $E_{n}^{\eta, s}(k_{y})$ is an even function of $k_{y}$, see Eq. (2).
Then if we interchange $n$ and $ n^{\prime}$ and  perform the $k_{y}$ integration  using Eqs. (7) and (8), we can write
the non-interacting density-density correlation function as 
\begin{align}
\Pi_{0}(q,\omega) & =\frac{1}{D_{0}}\underset{n,n^{\prime}}{\sum }%
J_{nn^{\prime}}(u)f(E_{n}^{\eta,s}) \notag \\
& \times\Big[(E_{n}^{\eta,s}-E_{n^{\prime}}^{\eta,s}+\hbar\omega+i\gamma
)^{-1}-(E_{n^{\prime}}^{\eta,s}-E_{n}^{\eta,s}+\hbar\omega+i\gamma )^{-1}\Big].
\label{9} 
\end{align}
The real and  imaginary parts of $\Pi_{0}(q,\omega)$ 
can be obtained   from the identity $1/(x\pm i\gamma)=(\wp/x)\mp
i\pi \delta(x)$ where $\wp$ denotes the principal value of $1/x$. The real
part of Eq. (9) reads 
\begin{equation}
\Pi_{1}(q,\omega)=\frac{1}{D_{0}}\underset{n,n^{\prime}}{\sum }%
J_{nn^{\prime}}(u)[I_{1}(\omega)+I_{1}(-\omega)],  \label{10}
\end{equation}
with 
\begin{equation}
I_{1}(\omega)=f(E_{n}^{\eta,s})/[E_{n}^{\eta,s}-E_{n^{\prime}}^{\eta
,s}+\hbar\omega],  \label{11}
\end{equation}
while the imaginary part 
is written as
\begin{equation}
\Pi_{2}(q,\omega)=\frac{\pi}{D_{0}}\underset{n,n^{\prime}}{\sum}%
J_{nn^{\prime}}(u)[I_{2}(\omega)-I_{2}(-\omega)],  \label{12}
\end{equation}
with 
\begin{equation}
I_{2}(\omega)=f(E_{n}^{\eta,s})\delta(\hbar\omega+E_{n^{\prime}}^{%
\eta,s}-E_{n}^{\eta,s}). \label{13}
\end{equation}

Equations (10)-(13) will be the starting point of our treatment of
 magnetoplasmons. Their form 
makes clear their even and odd symmetry with respect to $\omega$. 
These functions are the essential ingredients for theoretical considerations of
such diverse problems as high-frequency and steady-state transport, static
and dynamic screening, and correlation phenomena.

ii) {\it Limit $\omega =q=0$.}
The non-interacting density-density correlation function is obtained from Eq. (6)
in the static and long wavelength limit, $\omega =q=0$.
Thus Eq. (6) becomes
\begin{equation}
\Pi _{0}(0,0)=\frac{1}{D_{0}} \sum_{n,n^\prime,\pm}  
\frac{f(E_{n}^{\eta ,s})-f(E_{n^{\prime
}}^{\eta ,s})}{E_{n}^{\eta ,s}-E_{n^{\prime }}^{\eta ,s}},  \label{14}
\end{equation}
where the summation over +/- represents electrons/holes. With the zero-temperature limit, this turns into a series of delta functions, $\delta (E_{F}\pm
E_{n}^{\eta ,s})$ \cite{31,32}. Making the replacement $\delta
(E)=(\Gamma/\pi) /(E^{2}+\Gamma ^{2})$, we arrive at
\begin{equation}
\Pi _{0}(0,0)=\frac{1}{2\pi D_{0}}\sum_{n=0,\pm}^{\infty } 
\frac{(2-\delta _{0,n})\Gamma }{(E_{F}-E_{n}^{\eta
,s})^{2}+\Gamma ^{2}},  \label{15}
\end{equation}
where $\Gamma $ is the level width. Then the density-density correlation
function is proportional to the density of states at the Fermi energy, $\Pi
_{0}(0,0)=D(E_{F})$. At finite temperatures though it  
is given by  \cite{31,32} 
\begin{equation}
\Pi _{0}(0,0)=\int_{-\infty }^{+\infty }\left[ -\partial f(E)/
\partial E\right] D(E)dE. 
\label{16}
\end{equation}%
The density-density correlation function shows the lifting of the four-fold
degeneracy at $E_{F}=0$ (Dirac point) at 
zero temperature. At $T=0$ and  $E_{F}=0$ this
function vanishes  in the limit of zero SOC and $E_z$,
simply because 
it becomes the same as that of graphene 
at the Dirac point ($E_{F}=0$) with a completely filled valence band and completely empty
conduction band. The corresponding  carrier
density vanishes and implies that no intrinsic graphene
plasmons are possible (more generally, Dirac plasmons). This 
means that the screening is absent to  linear order except for the
renorrmalization of the dielectric constant term.
However, when the Fermi level
is away from $E_{F}=0$ or at nonzero temperature, the density-density
correlation function shows doubly degenerate spin ad valley splitting of the
Landau levels (LLs) and the linear screening is expected to become appropriate.
Moreover, these results can be reduced to those for gappless graphene derived
and discussed in Ref. \cite{32}  (see Fig. 1) in the limit of zero SOC\ and  $E_z$.

\begin{figure}[ht]
\begin{center}
\includegraphics[width=4.8 cm, height=5 cm]
{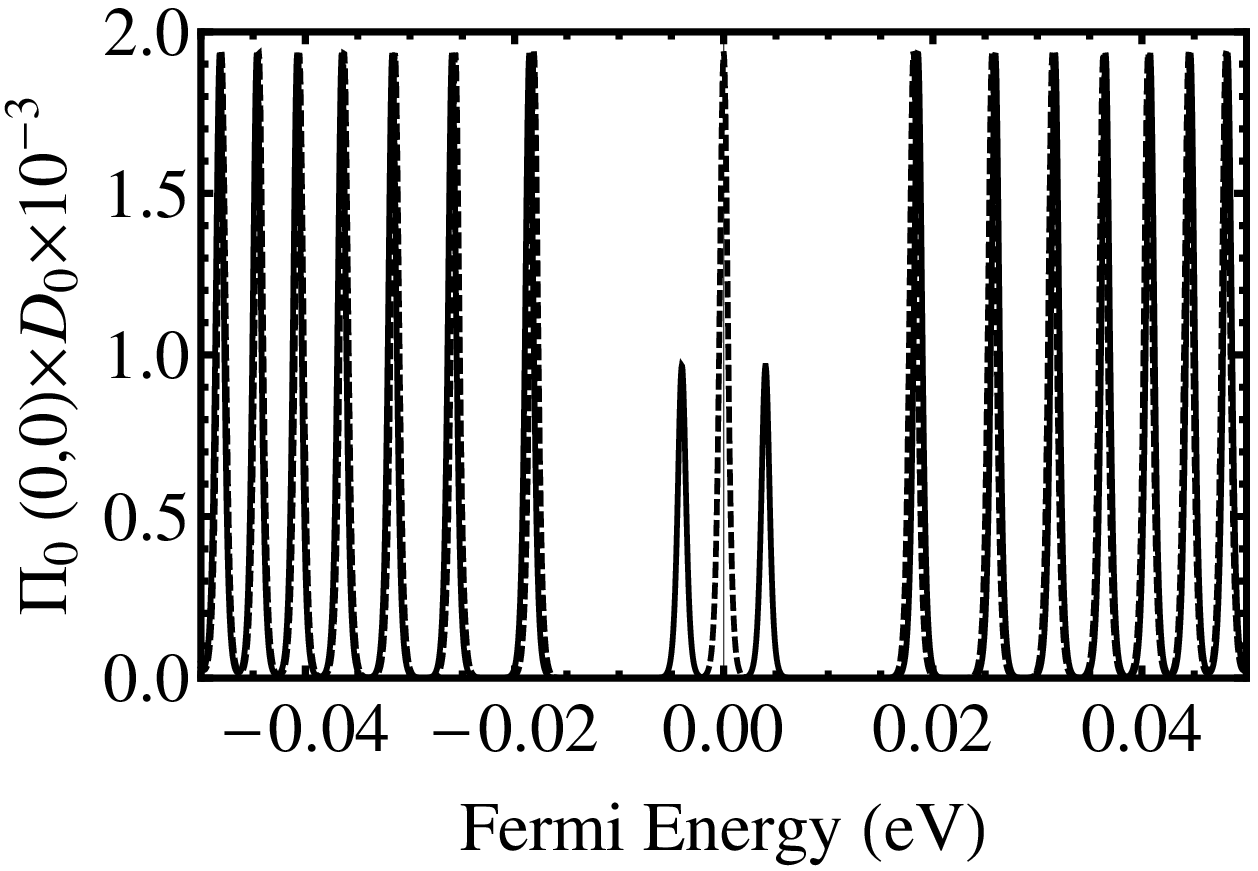} 
\hspace*{.4cm}
\includegraphics[width=4.8 cm, height=4.8 cm] 
{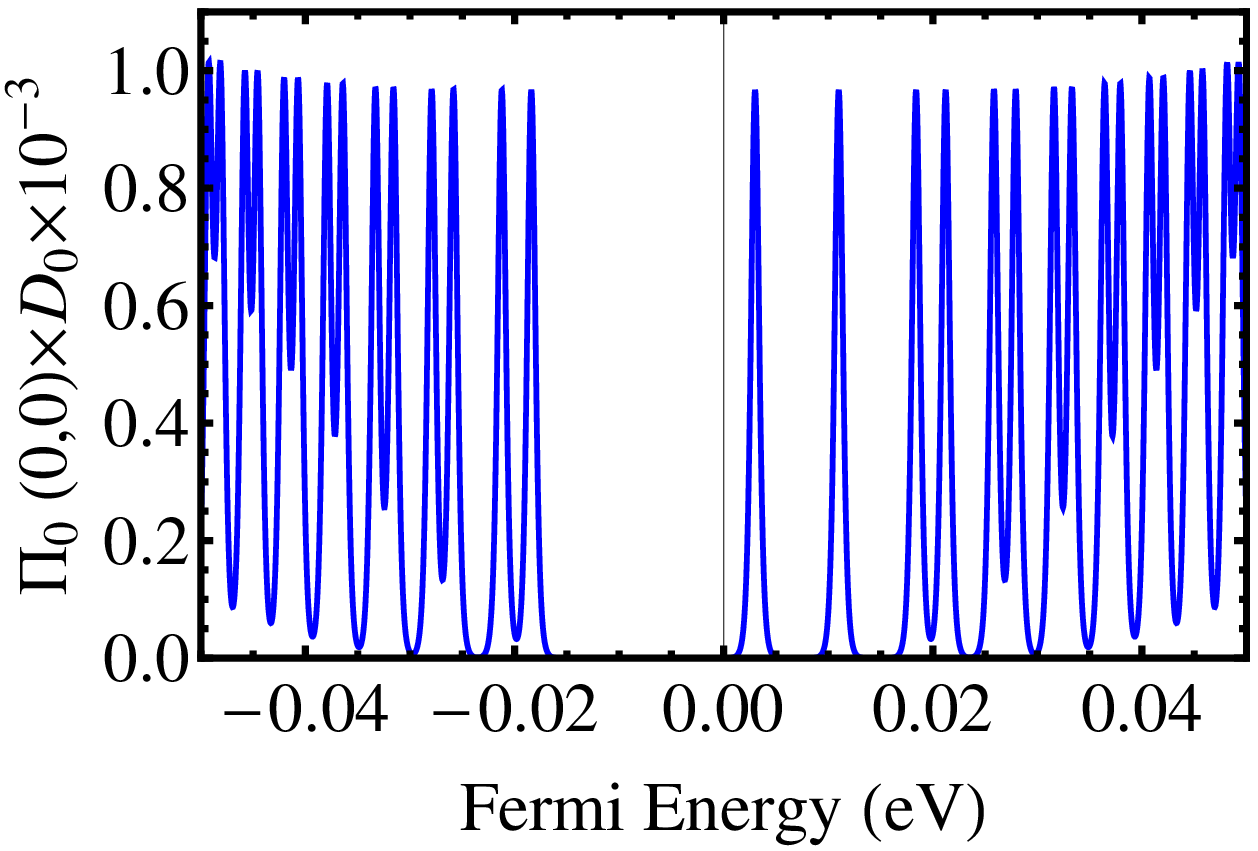} 
\hspace*{.4cm}
\includegraphics[width=4.8 cm, height=4.8 cm] 
{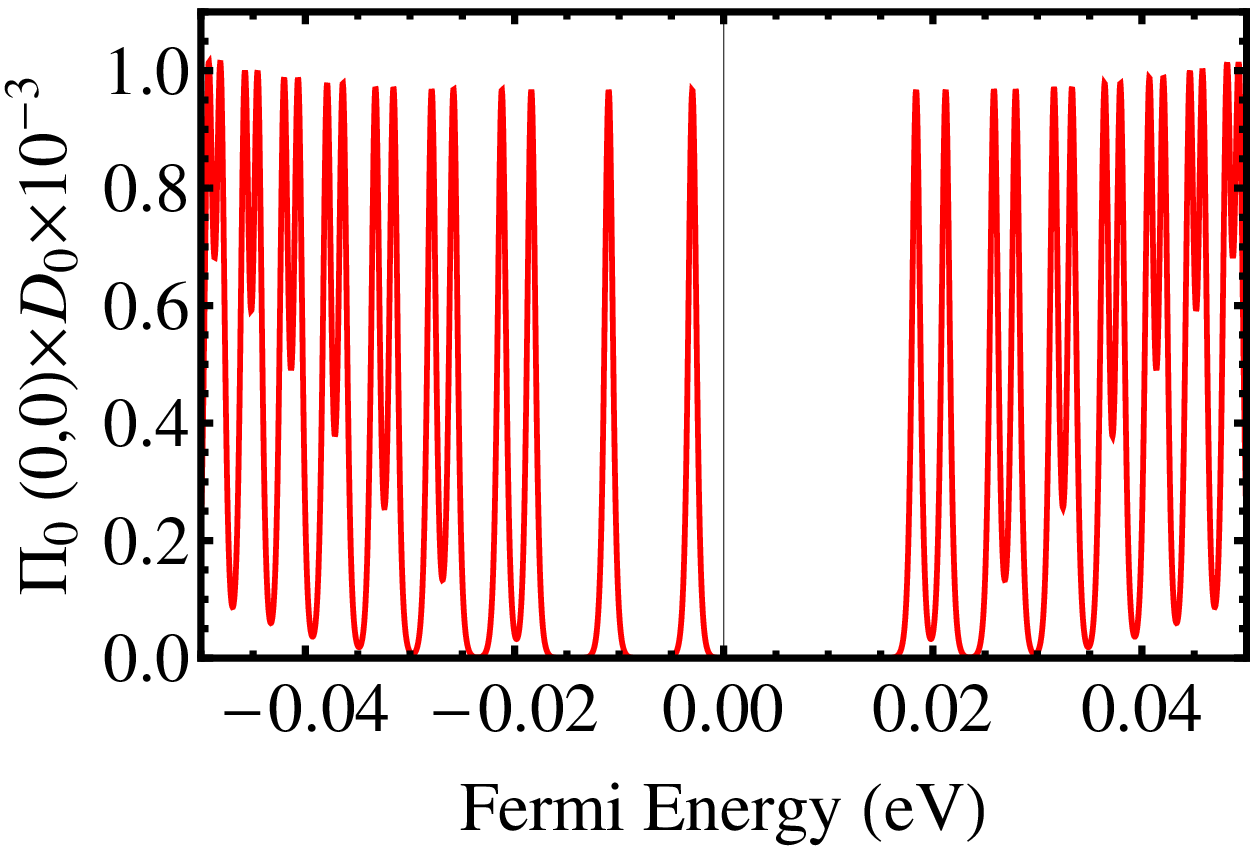}
\end{center}
\vspace*{-.7cm}
\caption{Static density-density correlation function
in  the long-wavelength limit $q\to 0$ versus  Fermi energy $E_F$. 
We vary the field $E_z$ and the SOC strength $\lambda$.
Left  panel: the  dashed and solid a curves are  for $V=\lambda =0$ and $V=0$ meV and $%
\lambda =4$ meV. The other two panels are for $V=7$ meV and $\lambda =4$ meV; the  middle panel is for the  $K$ valley and the the right one for the 
$K^{\prime }$ valley. The degeneracy of the LLs is lifted. }
\label{fig:1}
\end{figure}

We show  numerical results of Eq. (16) as a function of the Fermi energy
in Fig. 1. We find that the $n=0$ LL is split  into four levels and all
other LLs ($n>0$) into two. 
The valley degeneracy is lifted by the application of the field $E_z$ 
and the spin degeneracy by the SOC. This is
consistent with the eigenvalues given by Eq. (2). 
We use  $B$ = 1 Tesla, $T= 3$ K, and vary the field energy $V=2lE_z$ 
and the SOC strength. The left  panel is drawn  for $V=\lambda =0$ (dashed curves) and $V=0$ meV and $%
\lambda =4$ meV ( solid curves). The other two panels are for $V=7$ meV and $\lambda =4$ meV; the  middle panel is for the  $K$ valley and the the right one for the  $K^{\prime }$ valley.

In Fig. 1 
the SOC\ and   field split the LLs in two groups: in accordance with Eq. (2), $\eta,s={\pm }$,
we label them as  $+\equiv +,+\equiv-,-$ and $%
-\equiv -,+\equiv +,-$. Every $n\neq 0$ LL is doubly
degenerate in each group and  consists of a spin-up state from one valley
and a spin-down state from the other valley. The LL splitting
 between the two groups  is symmetric in the valence and
conduction band due to the symmetry in Eq. (2). 
The four-fold spin and valley degeneracy of the $n=0$ LL is lifted by the SOC and electric field energy. 

iii) {\it zero frequency}. The static limit $\omega\rightarrow0$ of Eq. (6) is obtained with the help of Eqs. (7)-(8). In this limit $\operatorname{Im}\Pi
_{0}(q,\omega)\rightarrow0$ and Eq. (6) gives  
\begin{equation}
\Pi_{0}(q,0)=\frac{1}{D_{0}}\sum_{n,n^{\prime}}
\frac{f(E_{n}^{\eta,s})-f(E_{n^{\prime}}^{\eta,s})}{E_{n}^{\eta,s}-E_{n^{\prime}}^{\eta,s}}\,J_{nn^{\prime}}(u). \label{17}
\end{equation}
We show numerical results for $\Pi_{0}(q,0)$  
as a function of the wave vector $q$ in Fig. 2. We use the parameters $B$ =
5 Tesla, $T$ = 10 K, and vary the  field energy $V=2lE_z$ and the SOC strength $\lambda$.  The  black curves
are for $V=\lambda=0$, the red ones for $V=0$ meV and $\lambda=4$ meV, and the blue curves
for $V=10$ meV and $\lambda=4$ meV. The solid and dashed curves pertain, respectively, to spin up and $K$ valley and to spin down and $K^{\prime}$ valley.

In the usual 2DEG the screening wave vector is independent of the carrier density but for
graphene or silicene it is proportional to the square root of the density \cite{16}.
First, 
in the limit of zero magnetic field $B$ the static correlation function
remains constant and equal to the electronic density of states 
up to the wave vector of $q=2k_{F}$; there are two contributions to it that stem from
intraband and interband plasmons, respectively. In the large momentum transfer
regime of Fig. 2, $q\sim 5$ ($10^8$ m$^{-1}$, the static screening for the  intraband case decreases linearly with $q$,
which is consistent with the case of gapless graphene in the limit of zero
(see Fig. 2 of Ref. \cite{16} and Fig. 4 of Ref. \cite{31}) and finite \cite{23} magnetic field.
There is no possibility of zero-energy plasmon excitations in the {\it intraband}
region (valence or conduction band).

\begin{figure}[ht]
\begin{center}
\vspace*{-.09 cm}
\includegraphics[width=0.5\columnwidth,clip]{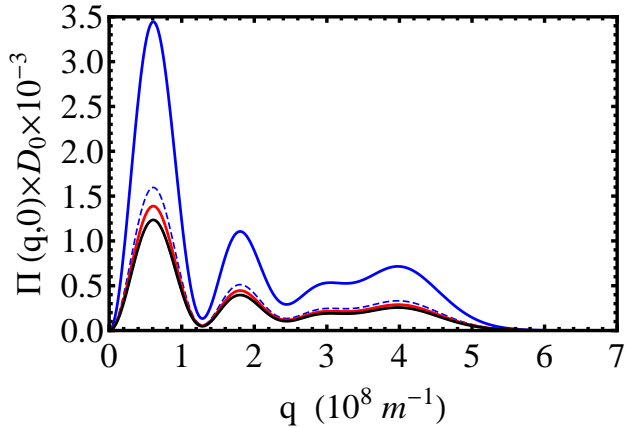} %
\end{center}
\vspace*{-.91 cm}
\caption{Static density-density correlation function
versus the wave vector $q$. We vary the electric field energy $V$ and the SOC strength $\lambda$.
Black curves: $V=\lambda=0$; red curves $V=0$ meV, $\lambda=4$ meV;  blue curves $V=10$ meV and $\lambda=4$ meV.
 The solid and dashed curves pertain, respectively, to spin up and spin down in the $K$ valley or spin down and spin up in the $K^{\prime}$ valley.
Here we cannot distinguish between the black (or red) solid and dashed curves
 as there is no spin or valley splitting for the chosen parameters.}
\label{fig:2}
\end{figure}

We find a similar behaviour for finite $B$  except in the small wave vector
limit. In contrast with its behaviour at $B=0$,  the static correlation function tends to zero as $\Pi_{0}(q\rightarrow0,0)\propto q^{2}$ for finite $B$  \cite{23}. 
This is due to the fact that the main contribution to it comes from the $q=0$ excitations in the vicinity
of $E_F$. Whereas at $B=0$  there are
$q\to 0$ excitations  whose energy  tends to zero, $E_F$ now lies in
a cyclotron gap between the highest occupied landau level $n_{F}$\ and
lowest unoccupied $n_{F}+1$. This   gap must be overcome by small-$q$
excitations, such that its spectral weight approaches zero. The static
correlation function also coincides with the density of states at $E_F$ because the latter vanishes for finite fields $B$  when 
$E_F$ is in the gap. Further, the oscillatory behaviour of the static correlation
function below  $2k_{F}$ is due to {\it intraband} transitions, whether $E_F$ is
in the valence or conduction band ($n_{F}%
+1,n_{F}$).

\section{Magnetoplasmons}

Magnetoplasmons are readily furnished by the singularities of the function $%
\Pi _{1}(q,\omega )$, from the roots of the longitudinal magnetoplasmon
dispersion relation obtained from Eq. (9) as
\begin{equation}
1-v_{c}(q)\Pi _{1}(q,\omega )=0,  \label{18}
\end{equation}%
along with the condition $\Pi _{2}(q,\omega )=0$ to ensure long-lived
excitations \cite{22,23,29,30}, which is in excellent agreement
with high-mobility graphene samples \cite{28}. 

For weak damping
the decay rate $\gamma$, determined by Eqs. (10) and (12), is given by Eq. (22) of Ref. \cite{30}.
Since we are primarily interested in the long-wavelength behavior of undamped magnetoplasmons,
described by $\gamma\propto\Pi _{2}(q,\omega )=0$, 
we treat them 
by solving Eq. (18).
With the help of Eq. (10) we find its roots  are obtained by solving 
\begin{equation}
1=\frac{ e^{2}}{kql^{2}}
\underset{n,n^{\prime }}{\sum }%
J_{nn^{\prime }}(u )\big[I_{1}(\omega )+I_{1}(-\omega )\big].  \label{19}
\end{equation}
Using Eq. (11) we can write
\begin{equation}
I_{1}(\omega )+I_{1}(-\omega )=\frac{2\Delta _{n,n^{\prime
}}^{\eta ,s}}{\hbar^2\omega^{2}-(\Delta _{n,n^{\prime }}^{\eta ,s})^{2}}\,f(E_{n}^{\eta ,s}),  \label{20}
\end{equation}
where $\Delta _{n,n^{\prime }}^{\eta ,s}=E_{n^{\prime }}^{\eta
,s}-E_{n}^{\eta ,s}$. Next we expand $J_{nn^{\prime }}(u)$ to lowest order
in its argument (low wave-number expansion). This amounts to considering
only the $n^{\prime }=n\pm 1$ terms in Eq. (19). The inter-Landau level
plasmon modes under consideration arise from neighbouring Landau levels, that is,
from $n^{\prime }=n\pm 1$.  Then using the  expansion \cite{33} $L_{n}^{^{l}}(u)= 
{\displaystyle\sum\limits_{m=0}^{n}}  
\frac{(-1)^{m}\,(n+l)!}{(l+m)!(n-m)!}\frac{u^{m}}{m!}$ for $l>0$ and
retaining only terms that are constant or linear in $u$ we get 
\begin{equation}
J_{n,n+1}(u )\rightarrow  nu G^{C}+(n+1)u G^{D},  \label{21}
\end{equation}
\begin{equation}
J_{n,n-1}(u )\rightarrow (n-1)u G^{C}+nu G^{D}.  \label{22}
\end{equation}%
Here $G^{C}=(1+r_\xi)/2$,  $G^{D}=(1-r_\xi)/2$, and $r_\xi=V_{\xi }/[\hslash ^{2}\omega
_{c}^{2}n+V_{\xi }^{2}]^{1/2}$. 
The factors $G^{C}$ and $G^{D}$ arise from the
normalization of the eigenstates and in the limit $\lambda=E_z=0$ 
are both equal to 1/2. 

To obtain the magnetoplasmon spectrum, we evaluate 
$\Delta
_{n,n^{\prime }}^{\eta s}$ for $n^{\prime }=n\pm 1$. We find 
\begin{equation}
\Delta _{n,n\pm 1}^{\eta ,s}=\pm\, \hbar\omega _{c}/(2\big[n+(V_{\xi}/\hbar\omega _{c})^{2}\big]^{1/2}). 
\end{equation}
Substitution of Eqs  (20)-(22) into Eq. (19)
yields 
\begin{equation}
1=\frac{e^{2}q}{\kappa }\sum_{n}\,\frac{|\Delta _{n,n\pm 1}^{\eta
,s}|}{\hslash ^{2}\omega ^{2}-\big(\Delta _{n,n\pm 1}^{\eta
,s}\big)^{2}}\,f(E_{n}^{\eta ,s}).  \label{24}
\end{equation}%
%
For inter-LL excitations near the Fermi energy $E_F$ we can approximate $n$ by $n_F$ in  $\Delta _{n,n\pm 1}^{\eta ,s}$, where  $n_{F}$  
is the LL index corresponding to   $E_{F}$. This gives 
\begin{align}
\hslash^{2}\omega^{2}&=(\hbar^2\omega _{c}^2/(4\big[n_F+(V_{\xi}/\hbar\omega _{c})^{2}\big])) \\*   \label{25}
&\times\Big[1+\frac{2e^{2}q \big[n_F+(V_{\xi}/\hbar\omega _{c})^{2}\big]^{1/2}}{\kappa \hbar\omega _{c} }\,\sum_{n}f(E_{n}^{\eta
,s})\Big].  \notag
\end{align}
With $E_{F}$ in the conduction band ($E_{F}^{2}=\hslash
^{2}\omega _{c}^{2}n_{F}+V_{\xi }^{2}$) 
 Eq. (25)  can be expressed as 
\begin{equation}
\omega ^{2}=\tilde{\omega}_{c}^{2}+\tilde{\omega}_{p}^{2},  \label{26}
\end{equation}
where  
\begin{equation}
\tilde{\omega}_{c}= 
\omega _{c}\big[\hbar\omega _{c}/(2( 
\hbar ^{2}\omega
_{c}^{2}n_{F}+V_{\xi }^{2})^{1/2})\big],  
\label{27}
\end{equation}
and 
\begin{equation}
\tilde{\omega}_{p}=\omega_{p}\big[v_F/
(\hbar ^{2}\omega _{c}^{2}n_{F}+V_{\xi }^{2})^{1/4}\big],  \label{28}
\end{equation}%
 with  $\omega_{p}= \big[e^{2}q\pi n_{c}/\kappa \big]^{1/2}$ and  $n_{c}=\underset{n}%
{\sum }
f(E_{n}^{\eta ,s})/(\pi \ell ^{2})$   the 2D carrier density.

It is interesting that Eq. (24) can be applied to the usual 2DEG for which  $\Delta _{n,n\pm 1}^{\eta ,s}=\pm \hslash \omega _{c}$. Then we obtain again Eq. (26) with 
$\tilde{\omega}_{c}$ and $\tilde{\omega}_{p}$ replaced, respectively,  by $\omega_{c}=eB/m$ and $\omega_{p}= \big[e^{2}q\pi n_{c}/\kappa \big]^{1/2}$, 
that is,  the well-known plasmon dispersion relation. One can also take the  limit   $V_\xi\to 0$ in Eqs. (24)-(28) and obtain the dispersion relation for monolayer graphene
\cite{23,28}. Then 
 $\Delta _{n,n\pm 1}^{\eta ,s}=\pm \hslash \omega _{c}/(2n^{1/2})$,  $\tilde{\omega}_{c}=\omega _{c}/(2\sqrt{n_F})$, and $\tilde{\omega}_{p}=(v_F/(2n_F^{1/4}\sqrt{
 \hslash \omega _{c}}))\,\omega_{p}.$  

In the limit of zero magnetic field, Eqs. (26)-(28)  reduce to recent work on silicene and germanene \cite{29,30}. Moreover, in
the limit of zero SOC and $E_z$, these relations are the same as that
for high-mobility graphene samples \cite{28} and could be applied to
highly doped graphene samples \cite{26,27} (for very large $n_F$ in Eq. (25). The $q$ dependence of Eq. (26), namely the $\sqrt{q}$ behaviour, is
common to 2D electron gas systems while the carrier density dependence is
 characteristic of the linear-in-$k$ dispersion relation of massless Dirac
fermions, for which $E_{F}=\hslash v\sqrt{\pi n_{c}}$. However, in the
present case we can see the effects of gapped silicene or germanene with
massive Dirac fermions and spin/valley splitting due to the combination of
the SOC\ and the  electric field $E_z$.

\section{Discussion of results}
A closer analytical examination of Eq. (26) shows the following aspects of the gapped
magnetoplasmon spectrum. 
If we set  $E_z=0$  in Eq.~(26) we 
 obtain a SOC-induced, small-gap   magnetoplasmon spectrum. Increasing $E_z$, we obtain a larger gap, splitting and
tuning of plasmons in silicene by combining it with the SOC. If we use
a   field  $E_z$ comparable to the SOC strength $\lambda$,\ then we expect
splitting of the magnetoplasmon modes due to the combination of the two in the quantity  $V+\eta s\lambda$. 
With further increase in $E_z$ , e.g.,   $E_z=2 \lambda$\  we can see an enhanced spin and valley
splitting of the magnetoplasmon spectrum due to the $V+\eta s\lambda$ factor in Eq. (21). 
Moreover, we note that the realization of topological phase
transitions could also be observed in the magnetoplasmon spectrum if we take
$E_z$ zero or less than $\lambda$ (spin-Hall regime), comparable
to $\lambda$ (semi-metallic regime),\ and then twice  $\lambda$ (valley-Hall regime). The  spin-Hall regime is a topological insulator while
the valley-Hall one is a band insulator. For $B\to 0$ these transitions are consistent with recent plasmon predictions \cite%
{29,30}. Below we consider the effect of an external
 field $B$ using the parameters
\cite{7,26,27,28,29,30}: $q=\pi /100$ nm$^{-1}$,
$v_F=0.5\times 10^{6}$ m/s, $\lambda = 4$ meV
for silicene ($43$ meV for germanene) on SiC with  dielectric constant $%
\kappa \simeq 4$ (different values do not qualitatively affect the 
results), and carrier density $n_{c}=0.5\times 10^{16}$ m$^{-2}$  
giving  $E_{F}=$ 41.3 meV.

The changes in the density of states  $D(E)$ discussed in Sec. III
and the approximations used to obtain the magnetoplasmons are reflected in the dependence of  $E_F$, e.g., on the mangetic field. At finite temperatures  the 2D carrier density $n_{c}$ is 
$n_{c}=\int_{-\infty }^{\infty }D(E)f(E)dE$, with  $D(E)$  for the LL spectrum 
obtained as
\begin{equation}
D(E)=\frac{1}{D_{0}}\left( \frac{1}{2}\sum_{\eta ,s}\delta
(E-E_{0}^{\eta ,s})+\sum_{n=1,\eta ,s}\delta (E-E_{n}^{\eta ,s})\right);
\label{29}
\end{equation}
the factor $1/2$ refers the fact the degeneracy of the zero LL is half that of the other LLs.
 Using Eq. (29) the result for $n_c$ becomes 
\begin{equation}
n_{c}=\frac{1}{D_{0}}\left( \frac{1}{2}\sum_{\eta ,s}f(E_{0}^{\eta
,s})+\sum_{n=1,\eta ,s}f(E_{n}^{\eta ,s})\right) .  \label{30}
\end{equation}%
\begin{figure}[ht]
\begin{center}
\vspace*{-.09 cm}
\includegraphics[width=0.5\columnwidth,clip]{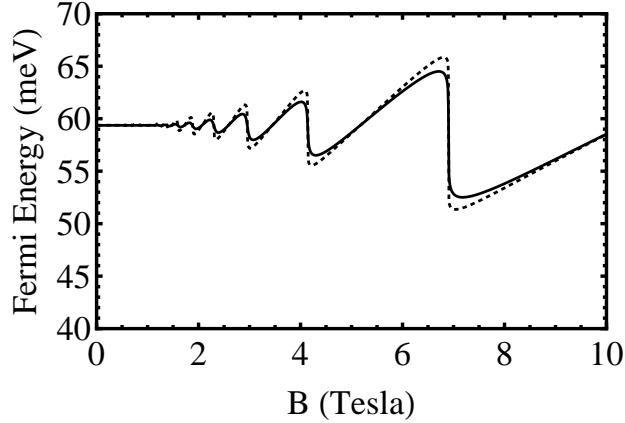} %
\end{center}
\vspace*{-.91 cm}
\caption{Fermi energy as a function of magnetic field
for fixed values  $V=10$ meV, $\lambda=4$ meV,
and $n_{c}=0.5\times 10^{16}$ m$^{-2}$. The
temperature is varied such that  $T=10$ K (solid) and  $T=5$ K (dotted).}
\label{fig:3}
\end{figure}

For fixed carrier density, this determines $E_F$   implicitly by solving numerically  Eq. (30).
We show the resulting $E_F$, as a function of
the  magnetic field $B$ in Fig. 3,  for 
$V=10$ meV, $\lambda=4$ meV, $T=10$ K, and $n_{c}=0.5\times 10^{16}$ m$^{-2}$.
$E_F$ remains  constant for 
low $B$ below 2T, that is,
in the limit of large $n$;  above this value we see the jumps as $E_F$ crosses the LLs.

\begin{figure}[ht]
\begin{center}
\includegraphics[width=5 cm, height=5 cm]  
{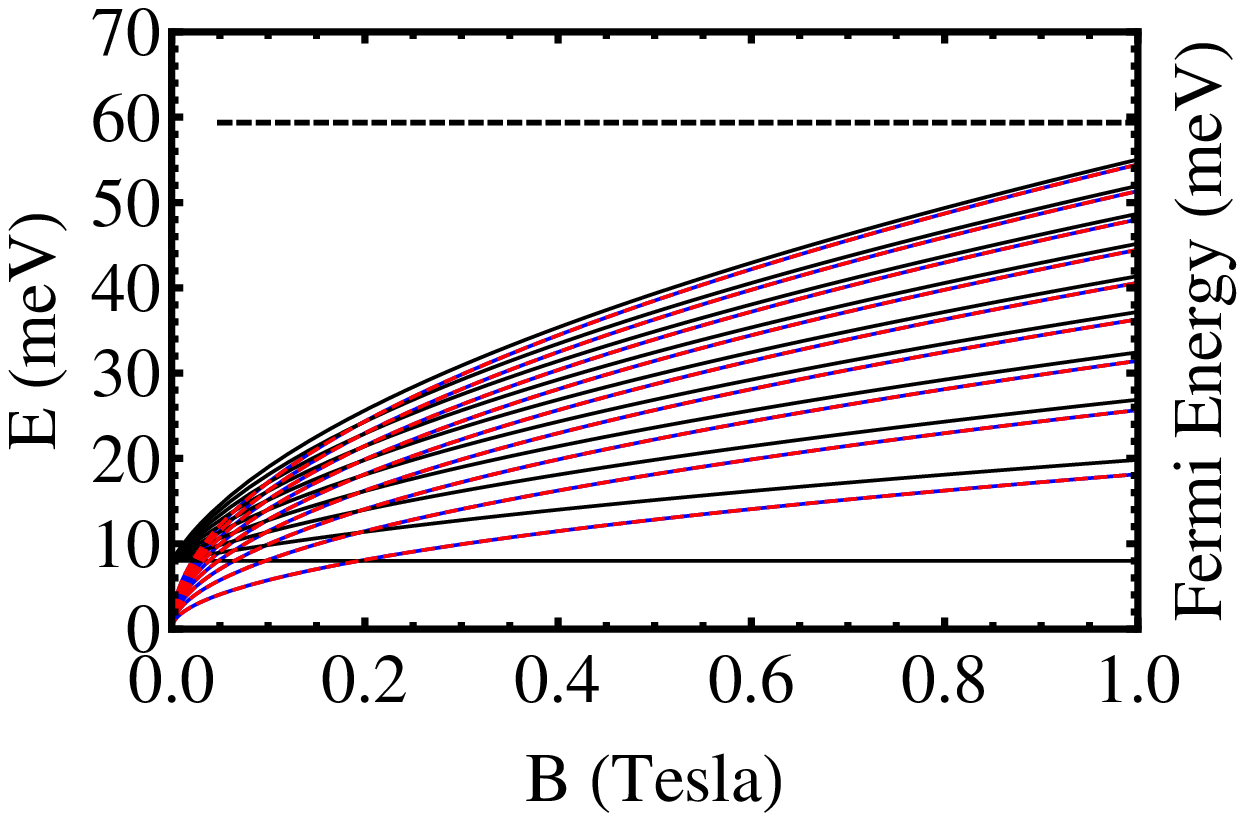} %
\hspace*{0.4cm}
\includegraphics [width=5 cm, height=5 cm] 
{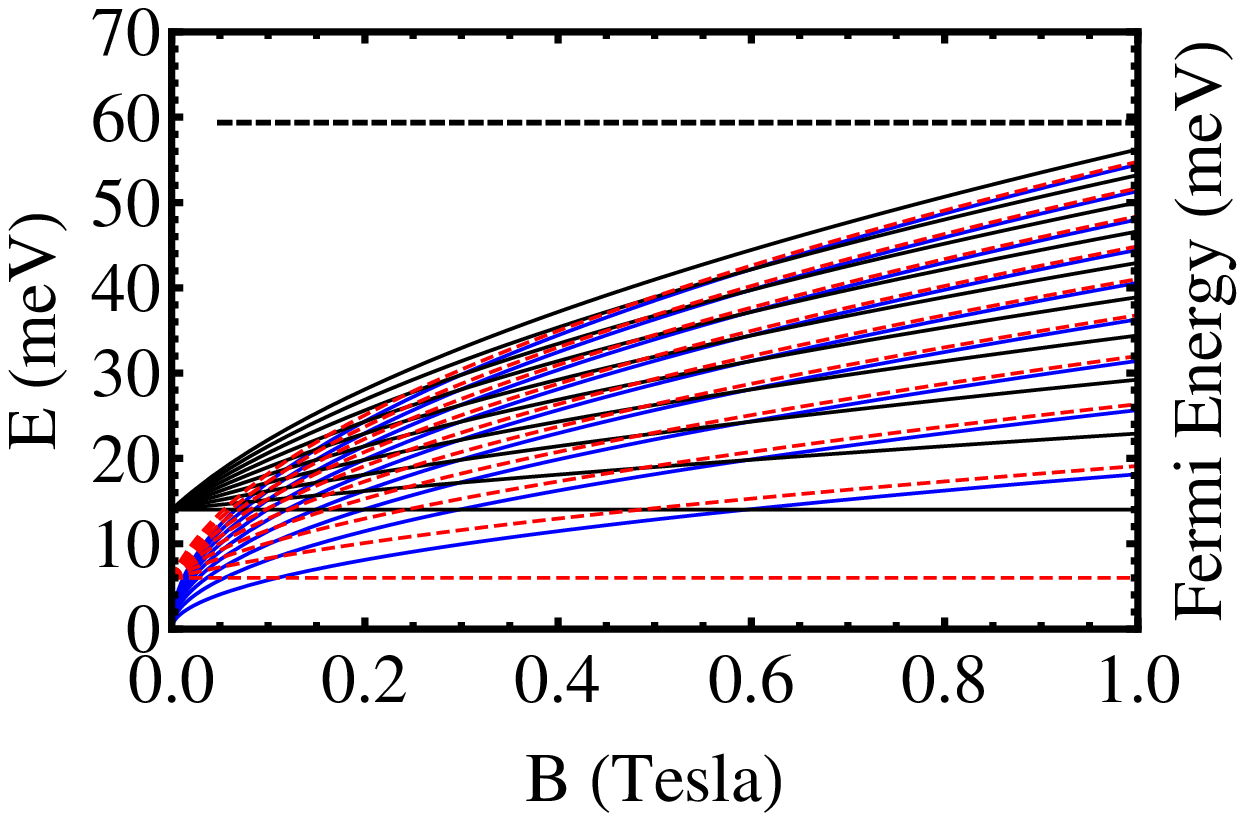} %
\hspace*{0.4cm}
\includegraphics [width=5 cm, height=5 cm] 
{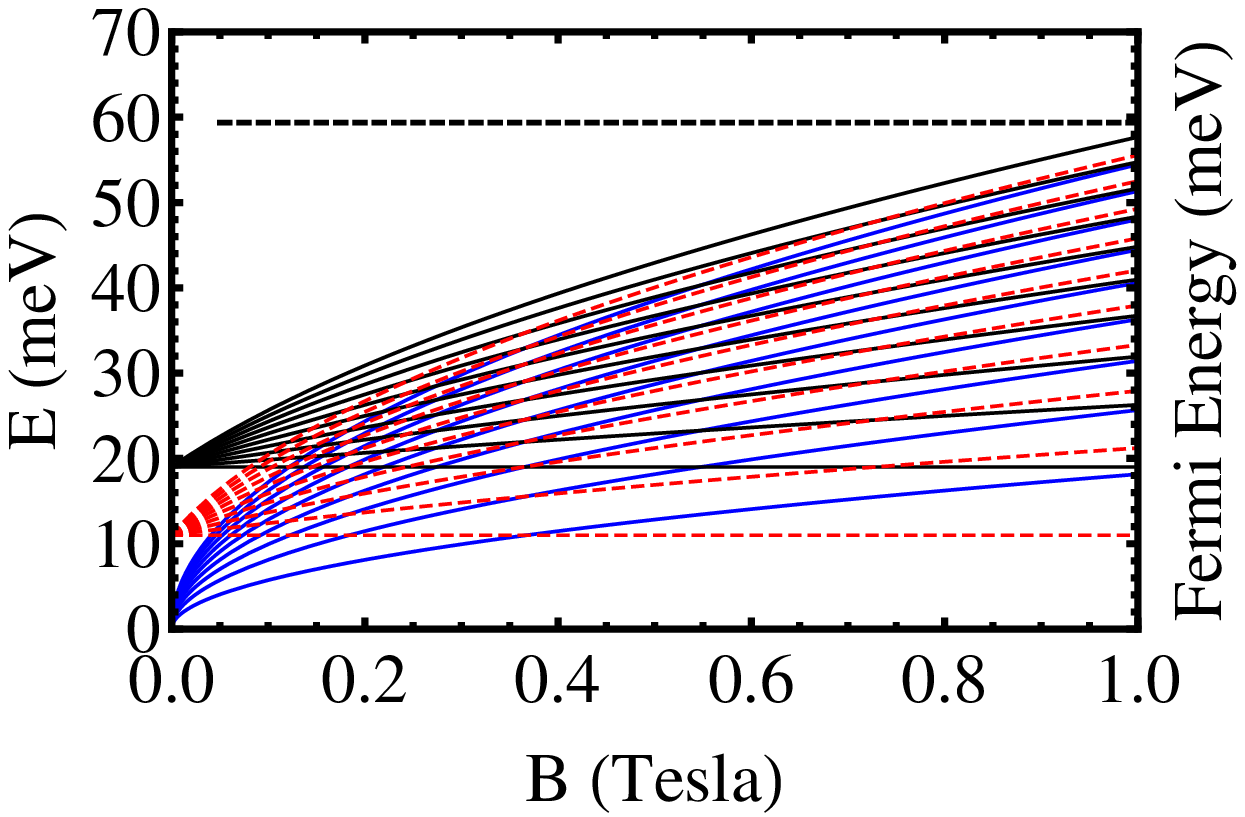}
\end{center}
\vspace{-0.7cm}
\caption{Band structure of silicene as a function of the magnetic
field $B$. The blue curves correspond to $\protect\lambda=V=0$ and the black and red dotted
ones to $E_{n}^{+}$ and $E_{n}^{-}$, respectively. The black dotted line
shows $E_F$ vs $B$ evaluated numerically using Eq. (30).
In the left panel we cannot distinguish between the blue and red dotted curves
because at $\protect\lambda=V=$ 4 meV the gap is zero
for the ($V-\protect\lambda$) curves.
In the middle panel we see a clear degree of spin and valley splitting for $\protect\lambda=$ 4 meV
and $V=$ 10 meV. The right panel,  
for $\protect\lambda=$ 4 meV and $V=$ 15 meV, shows a  significant
degree of spin and valley splitting by electrical tuning.}
\label{fig:4}
\end{figure}

\begin{figure}[ht]
\begin{center}
\includegraphics[width=5 cm, height=5 cm] 
{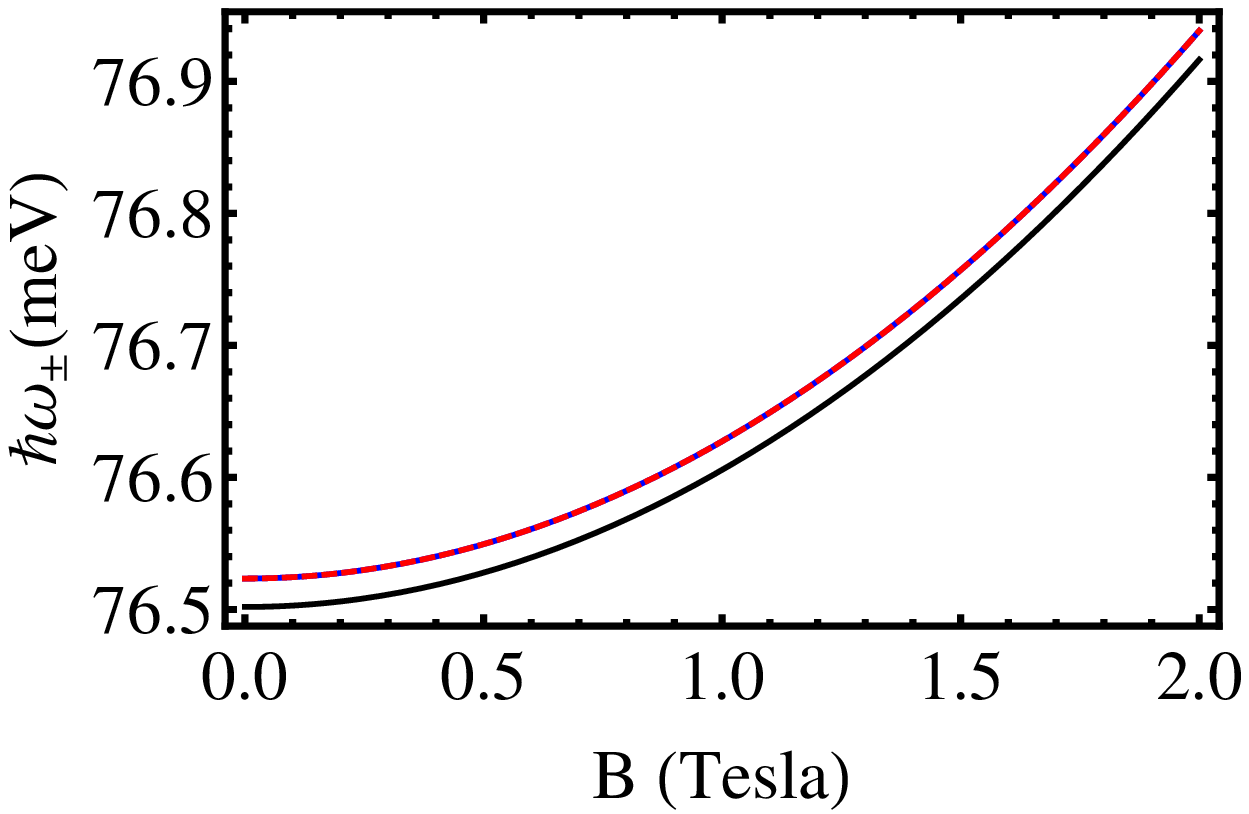} %
\hspace*{.4cm}
\includegraphics[width=5 cm, height=5 cm] 
{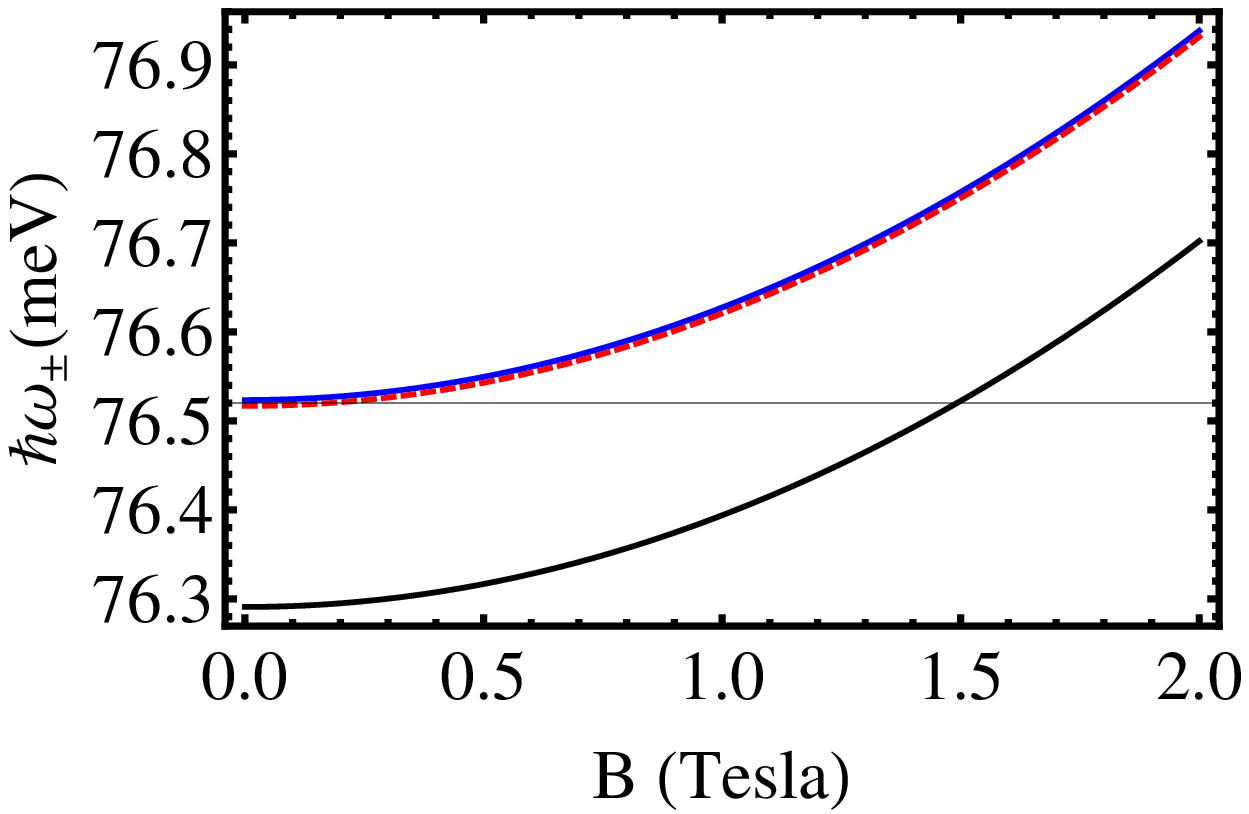} %
\hspace*{.4cm}
\includegraphics[width=5 cm, height=5 cm] 
{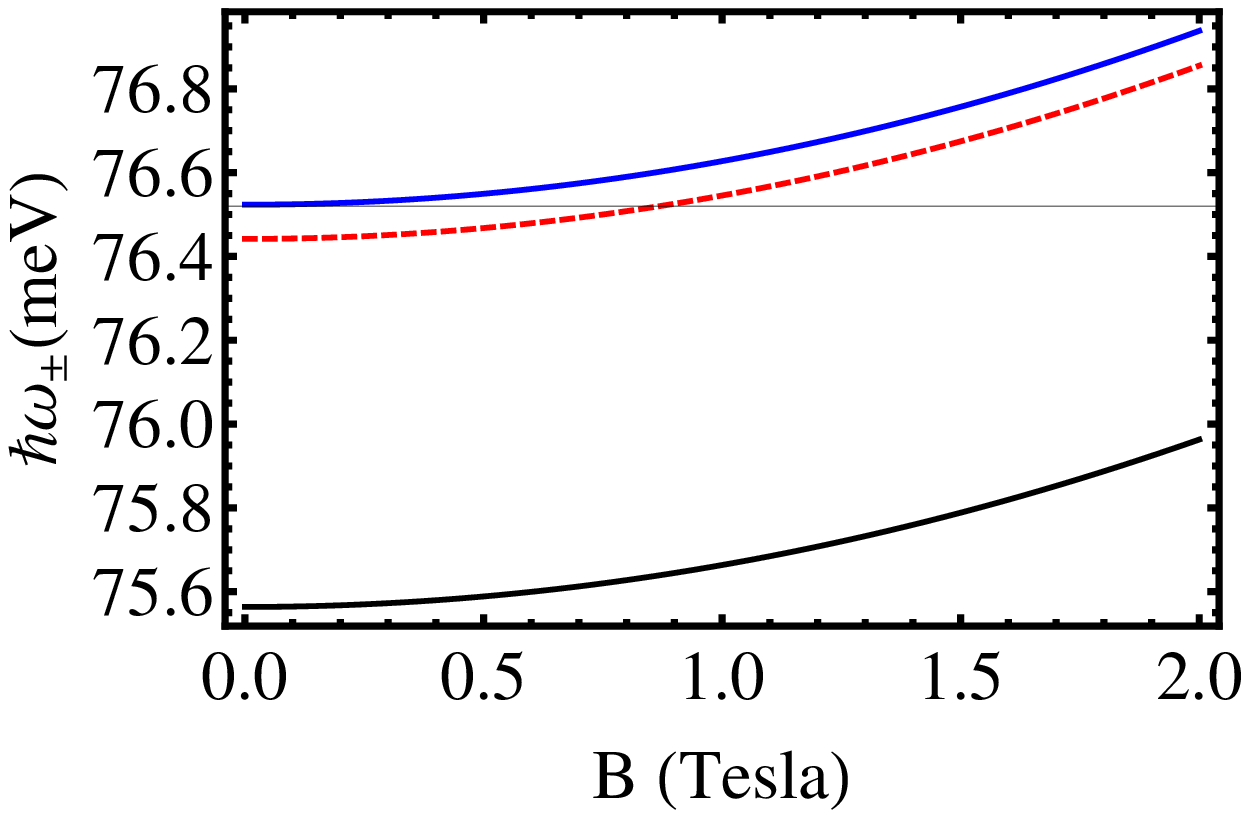}
\end{center}
\vspace{-0.7cm}
\caption{Magnetopasmons as a function of the magnetic field $B$ for a
fixed $E_F= 41.3$ meV. Blue curves  correspond to $\protect\lambda=V=0$.%
 Black and red dotted curves  represent $\hslash\protect\omega_{+}$ and $\hslash%
\protect\omega_{-}$, respectively. In the left panel, with $\protect\lambda=V=$ 4 meV, we
cannot distinguish between the red and blue  dotted curves  because at equal amount
of $V$ and $\protect\lambda$ energies, the gap is zero for the ($V-\protect\lambda$)
curves. The middle panel, for $\protect\lambda=$ 4 meV and $V=$ 10 meV, shows a clear
signature of spin and valley splitting. The right panel, for $\protect\lambda=$ 4
meV and $V=$ 15 meV, shows a significant  spin and valley
splitting by electrical tuning. The colour code is the same as in Fig. 4.}
\label{fig:5}
\end{figure}

\begin{figure}[ht]
\begin{center}
\includegraphics[width=5 cm, height=5 cm]  
{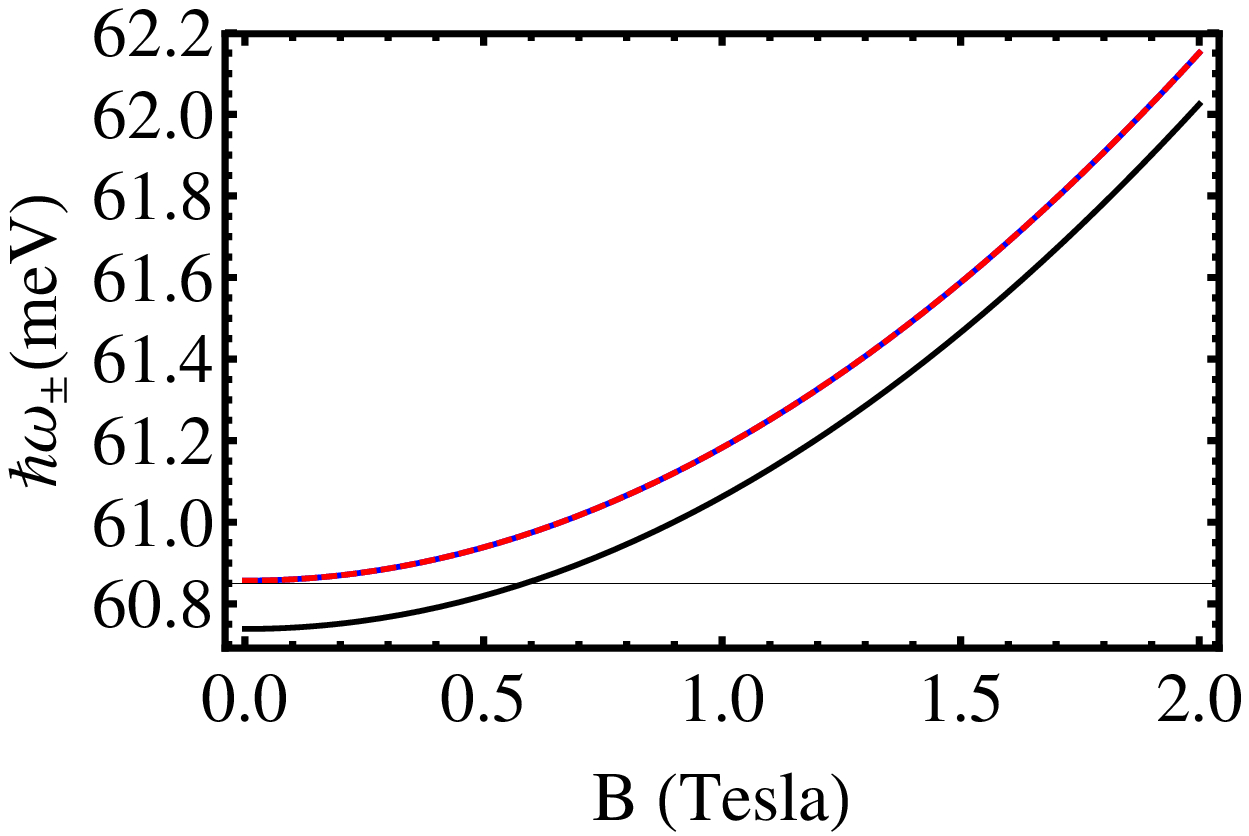} %
\hspace*{.4cm}
\includegraphics[width=5 cm, height=5  cm] 
{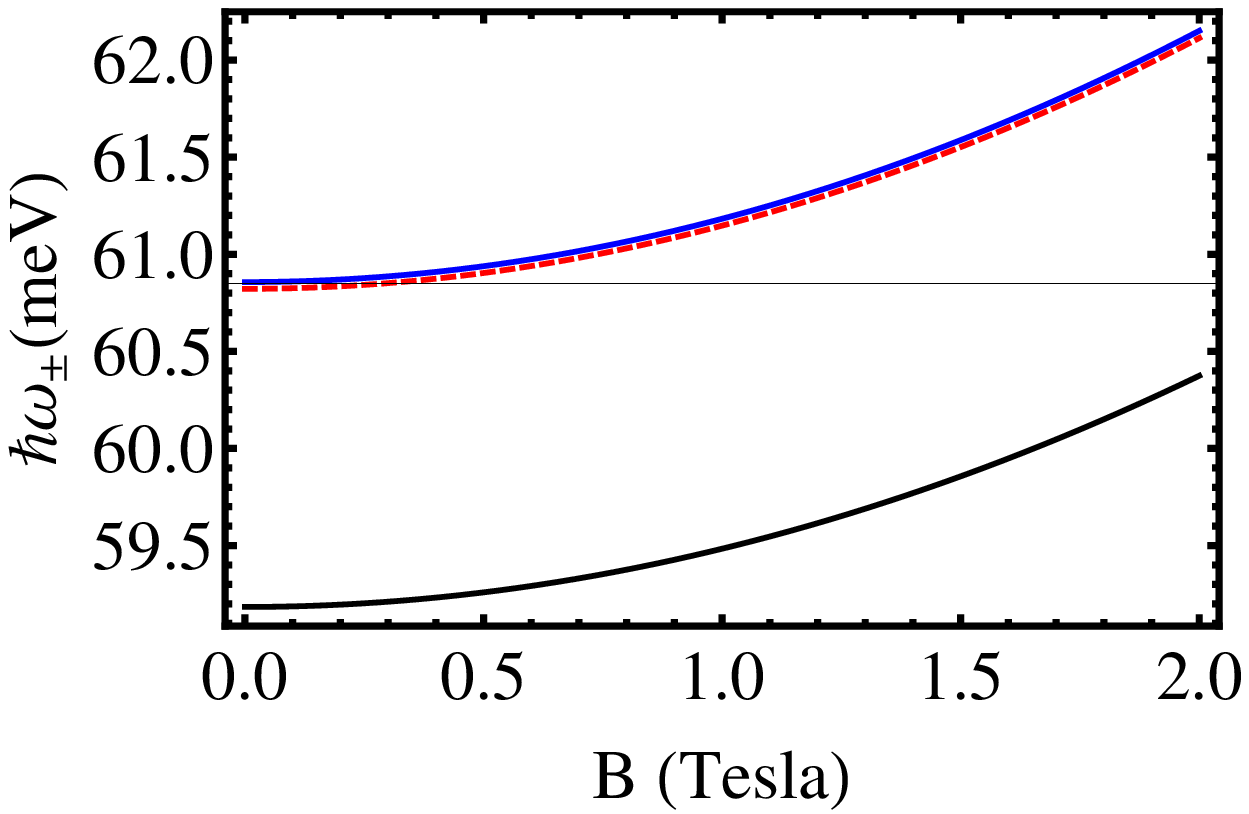}%
\hspace*{.4cm}
\includegraphics[width=5 cm, height=5  cm] 
{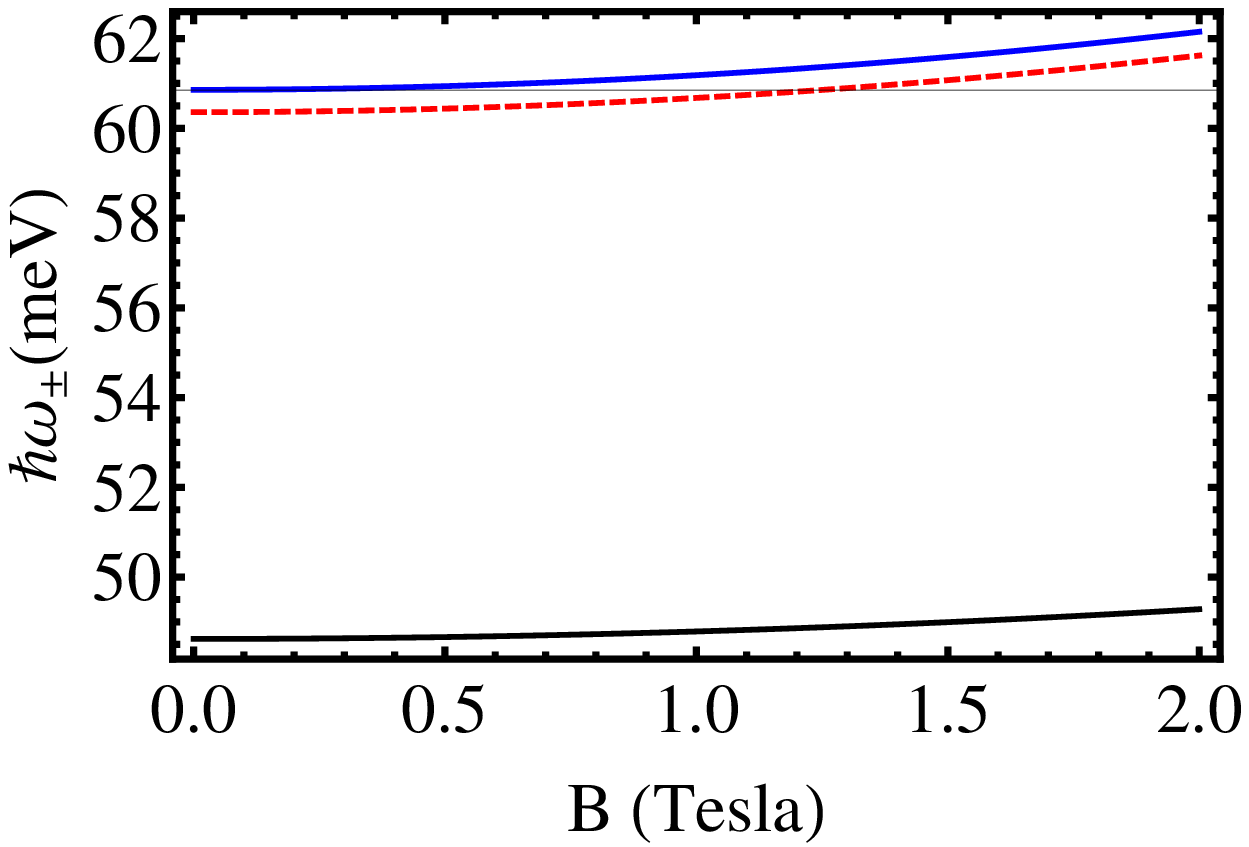}
\end{center}
\vspace{-0.7cm}

\caption{Magnetopasmons as a function of the magnetic field $B$ for a
fixed  $E_F=26$ meV. The parameters and curve marking are the same as in Fig. 5.}
\label{fig:6}
\end{figure}


We present the eigenvalues given by Eq. (2) as a function of the 
field $B$ for fixed values of $\lambda$ and $E_z$ in Fig. 4. We also include
the $E_F$ versus  field $B$ curve (dotted line) for
comparison and further discussion. We find the following:
(i) In the limit of $\lambda=\ell E_z=0$ (blue curves),
we obtain the $\sqrt{B}$ dependence of the LL energies. In contrast, for finite $\lambda$ 
 and variable $E_z$ (black and red dotted curves), the energies of the lower LLs grow linearly
with $B$ rather than with $\sqrt{B}$ because of the 
massive 
Dirac fermions in silicene or germanene. 
(ii) The combination of the  field energy $V=2\ell E_z$ and $\lambda$ splits the LLs in two groups 
designated as $E_{n}^{\pm}$, with  $E_{n}^{+}\equiv E_{n}^{+,+}=E_{n}^{-,-}$ and 
$E_{n}^{-}\equiv E_{n}^{-,+}=E_{n}^{+,-}$. (iii) The energies of the two groups of LLs in the valence or conduction band have not only different slopes versus $B$ but also shift rigidly for $B\to 0$ due to the finite band gap either by $\lambda$ or
by the field $E_z$. However, every $n\neq 0$ LL is still doubly
degenerate in each group, consisting of a spin-up state from one valley
and a spin-down state from the other valley. A crossing
occurs between the two groups, which is symmetric in the valence and
conduction band due to the symmetry in Eq. (2).

In Fig. 5 we show the magnetoplasmon spectrum as a function of the  field $B$ for
fixed   $E_{F}$ = 41.3 meV. For comparison with graphene experiments \cite{28}, we
show numerical results using Eq. (26) for  $\lambda=V=0$ (blue curve).
These results agree well with Eq. (1) and Fig. 2 of Ref. \cite{28}, exhibiting dependence on $\sqrt{B}$, if we replace $v_F=0.5\times 10^6$ m/s by its value in graphene $v_F=1\times 10^6$ m/s.
In the middle panel,  for finite $\lambda$ and $V=2lE_z$, we found two curves, the red dotted ($V-\lambda$) and black ($V+\lambda$)
  showing a spin and valley splitting.
The red dotted curve is the same as the blue one and we can't distinguish between the
two because the gap for the red dotted line vanishes due to $\lambda-2lE_z=0$.
As the gap due to $\lambda$ and $V$ is small and we are in a highly doped regime,
we can see a split between the red dotted and black curve for two magnetoplamon modes $\hslash \omega
_{\pm }$ defined as $\hslash \omega _{+}=\hslash \omega _{+,+}=\hslash
\omega _{-,-}$ and $\hslash \omega _{-}=\hslash \omega _{-,+}=\hslash \omega
_{+,-}$. Increasing  $V=2lE_z=10$ meV, we see
an enhanced splitting between red dotted and black curves for fixed $\lambda=4$ meV
(middle panel). Here the blue and red dotted curves are weakly separated as the gap vanishes for the
blue and   red dotted curves $V-\lambda=6$ meV. With further increase in $V$, $V=15$ meV, we
obtain a further enhanced splitting between the black and red dotted curves of the
magnetoplasmon modes as shown in the right panel for  fixed $\lambda=4$ meV.
We also note that the blue and red dotted lines are well separated as gap is zero for the
blue and $V-\lambda=$ 11 meV for the red dotted curve.

We contrast our results 
with those of recent graphene experiments on
high-mobility or weakly doped samples \cite{28}, in the limit  $\lambda=2\ell E_z=0$, by further decreasing
the Fermi energy close to the Dirac point. First, we show the magnetoplasmon
spectrum as a function of the field $B$ for   $E_{F} = 26$ meV in Fig. 6.
We found a clear splitting between the black and red dotted curves for
$\lambda=2lE_z=4$ meV (left panel) as 
in the left panel of Fig. 5. As $\lambda$ and $E_z$ are 
small and we are in a weakly doped regime, we can see a strong splitting for
the magnetoplamon modes $\hslash \omega _{\pm }$. Again here we cannot
distinguish between the red dotted and blue curves for the same reason as
 in Fig. 5. Upon increasing  $V$, e.g. to $V=10$ meV, we see  a large splitting between
the red dotted and black curves for fixed $\lambda =
4$ meV (middle panel). We can weakly distinguish between the blue and
red dotted curves here  since the gap is $V-\lambda=6$ meV for the red
dotted curve and zero for the blue one.
With further increase in $V$, $V=15$ meV, we
obtain a significant splitting between the red dotted and black curves
of the magnetoplasmon modes as shown in the right panel.
Here we also note that the blue and red dotted curves are well separated compared
to those in the right panel of the Fig. 5.
Again the results exhibit a square-root dependence on  B and agree
with recent graphene theory \cite{21,22,23,24,25}
and experiments \cite{28} in the limit  $\lambda=2\ell E_z=0$ provided we use $v_F=1\times 10^6$ m/s.

 The experimentally observed \cite{28}, $\sqrt{B}$ dependence of the spectrum referred to above, in the limit $\lambda=V\to 0$, applies to  high-mobility 
{\it weakly-doped} graphene  samples, cf. Fig. 6. 
For {\it highly-doped} samples \cite{26,27}  though that involve values
of $E_F\gg\lambda,V$, with $E_F$ of the order of  $200-300$ meV, the magnetoplasmon gaps and spilttitings reported above will be very difficult to achieve as they would require unrealistically high values of $V$. Notice though that our analysis for silicene  also holds for germanene,  a monolayer of
germanium, which has a much stronger SOC  than  silicene \cite{7,8}, $\lambda\approx 43$ meV. In both cases the predicted gaps and spilttings are sizeable for $E_F$ not too far from the Dirac point.

Another feature of our results is the magnetoplasmon gaps. Although not yet experimentally confirmed,  the SOC induced gap in silicene is about 1.55 meV [6,7] and is expected to be observed using  existing experimental techniques. In the present work on electrically tunable magnetoplasmons in silicene, we have obtained a gap of about 1 meV in Fig. 5 and 12 meV in Fig. 6 tuned by an external perpendicular electric field, which can be  further enhanced by increasing this electric field and lowering the Fermi energy of the system close to the Dirac point. We believe that this gap can be observed in experiments similar to those on high-mobility graphene samples studying magnetoplasmons [28]. 

A possible extension of our work 
would be to include an in-plane electric field 
and study  magneto-electric-plasmons. One could then use the eigenfunctions and eigenvalues derived in Ref. \cite{per} for $E_z=\lambda=0$ as a starting point.

\section{Summary}

We  showed electrically tunable effects in the
magnetoplasmon spectrum of silicene and germanene due to the spin and valley polarization.
Employing the RPA and including the effects of SOC\ and of an external electric field, we found a significant splitting of the magnetoplasmon spectrum. Our results agree well with graphene theory and experiments in the limit of
vanishing SOC and electric field provided $E_F$ is not too far from the Dirac point,  that is, for {\it weakly-doped} graphene  samples \cite{28}, if we use graphene's value for $v_F$. We expect that experimental studies of these novel phenomena in silicene, similar to those of Ref. [28], will be  very appropriate 
 since they directly bear on the many-body properties of silicene or germanene. Encouraging in this direction is the very recently reported local formation of high-buckled silicene nanosheets  realized on a MoS$_2$ surface [35].
 
 Electronic address: $^{\dag}$m.tahir06@alumni.imperial.ac.uk
 
 \appendix
\numberwithin{equation}{section}
\section{}
Below we outline the derivation of Eq. (8).
 The factor $J_{\alpha,\alpha^{\prime}}(u)$ in Eq. (7) 
 is given by 
\begin{equation}
J_{\alpha,\alpha^{\prime}}(u)=\left\langle \alpha^{\prime}\right\vert
\left\vert e^{-i\mathbf{q\cdot r}}\right\vert \left\vert \alpha\right\rangle ^{2}%
=\left\langle \alpha^{\prime}\right\vert \left\vert e^{-i\mathbf{q\cdot r}}\right\vert
\left\vert \alpha\right\rangle \times\left\langle \alpha\right\vert \left\vert
e^{i\mathbf{q\cdot r}}\right\vert \left\vert \alpha^{\prime}\right\rangle, \label{A.1}%
\end{equation}
where $\left\vert\alpha\right\rangle =\left\vert s,n,\eta
,k_{y}\right\rangle$. Using 
the eigenfunctions given by Eq. (3) 
 Eq. (A.1) takes the form 
\begin{equation}
\left\langle \alpha^{\prime}\right\vert \left\vert e^{-i\mathbf{q\cdot r}}\right\vert
\left\vert \alpha\right\rangle =\frac{1}{L_{y}}\sum_{n,\eta,s}%
{\displaystyle\int}
dye^{i\left(  k_{y}-k_{y}^{\prime}-q_{y}\right)  y} 
{\displaystyle\int\limits_{-\infty}^{\infty}}
dx\left(
\begin{array}
[c]{c}%
-iC_{n}^{\eta,s}\phi_{n-1}(\bar{x})\\
D_{n}^{\eta,s}\phi_{n}(\bar{x})
\end{array}
\right) ^{T}e^{-iq_{x}x}\left(
\begin{array}
[c]{c}%
-iC_{n}^{\eta,s}\phi_{n-1}(\bar{x})\\
D_{n}^{\eta,s}\phi_{n}(\bar{x})
\end{array}
\right)  \label{A.2},%
\end{equation}
where the superscript $T$ denotes the transpose of the column vector. With the help of the identity
$(1/L_y)
\int dy\,e^{i\left(  k_{y}-k_{y}^{\prime}-q_{y}\right)  y}=
\delta_{k_{y}%
^{\prime},k_{y}-q_{_{y}}}$ we can write Eq. (A.2) as 
%
\begin{equation}
\left\langle \alpha^{\prime}\right\vert \left\vert e^{-i\mathbf{q\cdot r}}\right\vert
\left\vert \alpha\right\rangle =\delta_{k_{y}^{\prime},k_{y}-q_{_{y}}}%
\sum_{n,\eta,s}\Big[F_{n^{\prime},n}\left(  -q_{x},k_{y}-q_{y},k_{y}\right)
+F_{n^{\prime}-1,n-1}\left(  -q_{x},k_{y}-q_{y},k_{y}\right)\Big]  \label{A.4}.%
\end{equation}
Similarly, 
\begin{equation}
\left\langle \alpha\right\vert \left\vert e^{i\mathbf{q\cdot r}}\right\vert \left\vert
\alpha^{\prime}\right\rangle =\delta_{k_{y}^{\prime},k_{y}-q_{_{y}}}%
\sum_{n,\eta,s}\Big[F_{n,n^{\prime}}\left(  q_{x},k_{y},k_{y}-q_{y}\right)
+F_{n-1,n^{\prime}-1}\left(  q_{x},k_{y},k_{y}-q_{y}\right) \Big ]. \label{A.5}%
\end{equation}
Combining Eqs. (A.3) and (A.4), we arrive at%
\begin{align}
\nonumber
J_{n,n^{\prime}}(u)  &  =\delta_{k_{y}^{\prime},k_{y}-q_{_{y}}}\sum_{n,\eta
,s}\Big[F_{n^{\prime},n}\left(  -q_{x},k_{y}-q_{y},k_{y}\right)  \times
F_{n,n^{\prime}}\left(  q_{x},k_{y},k_{y}-q_{y}\right) \label{A.6}\\
&  \hspace*{2.3cm}+F_{n^{\prime}-1,n-1}\left(  -q_{x},k_{y}-q_{y},k_{y}\right)  \times
F_{n-1,n^{\prime}-1}\left(  q_{x},k_{y},k_{y}-q_{y}\right) \Big ].
\end{align}
Now we proceed with the evaluation of $F_{n^{\prime}n}\left(  -q_{x}%
,k_{y}-q_{y},k_{y}\right)  $. Using the explicit form of the harmonic oscillator functions $\phi_{n}(\bar{x})$ we have 

\begin{align}
\nonumber
 &F_{n^{\prime}n}\left(  -q_{x},k_{y}-q_{y},k_{y}\right) 
 =\frac{[D_{n^{\prime}}^{\eta,s}D_{n}^{\eta,s}]}{l^2\sqrt{\pi2^{n}%
2^{n^{\prime}}n!n^{\prime}!}}
{\displaystyle\int\limits_{-\infty}^{\infty}}
dX\text{ }
e^{  -\left( X+l\left(  k_{y}%
-q_{y}\right)  \right)  ^{2}/2}  
\\
& \hspace*{3.9cm} \times H_{n^{\prime}}\left(  X+l\left(  k_{y}-q_{y}\right)  \right)
e^{ -iq_{x}x}\, 
e^{ -
\left(  X+lk_{y}\right) ^{2}/2} 
\,H_{n}\left(
X+lk_{y}\right), 
\end{align}
where $X=x/l$.  Making the change 
$Y=X+lk_{y}+l\left(  -q_{y}%
+iq_{x}\right)/2  $ in Eq. (A.6) yields
\begin{align}
\nonumber
F_{n^{\prime}n}\left(  -q_{x},k_{y}-q_{y},k_{y}\right)   &  {\small =}%
\frac{[D_{n^{\prime}}^{\eta,s}D_{n}^{\eta,s}]}{\sqrt{\pi2^{n}2^{n^{\prime}%
}n!n^{\prime}!}}\,e^{ -u^{2}} 
e^{i l^{2}q_{x}\left(  -q_{y}+2k_{y}\right)/2 } \label{A.8}\\
&  \hspace*{-0.1cm}\times%
{\displaystyle\int\limits_{-\infty}^{\infty}}
dY\,e^{ -Y^{2}} 
H_{n^{\prime}}\left(  Y-l\left(
q_{y}+iq_{x}\right)/2  \right)  H_{n}\left(  Y-l\left(  q_{y}%
-iq_{x}\right)/2  \right), 
\end{align}
where $u=l^{2}{q}^{2}/2$. The integral over $Y$ is tabulated in Ref.
34, 
pp. 838 \#7.377. 
The result for $n\leq
n^{\prime}$ is
\begin{align}
\nonumber
F_{n^{\prime}n}\left(  -q_{x},k_{y}-q_{y},k_{y}\right)   &  =\left( n!/n^{\prime}!\right) ^{1/2}
e^{ -u/2+il^{2}q_{x}\left(  -q_{y}+2k_{y}\right)/2  }\label{A.9}\\
&  \times\left[ l\left(  q_{y}-iq_{x}\right)/\sqrt{2}  \right]
^{n^{\prime}-n}L_{n}^{n^{\prime}-n}\left(  u\right). 
\end{align}
For $n^{\prime}\leq n$, the result is given by Eq. (A.8) with $n$ and $n'$ interchanged.
%
Using Eqs. (A.5) and 
(A.8) we arrive at Eq. (7),
\begin{equation}
J_{n,n^{\prime}}(u)=\ \left\vert\left\langle \alpha^{\prime}\right\vert 
e^{-i\mathbf{q\cdot r}} \left\vert \alpha\right\rangle\right\vert ^{2}=\delta_{k_{y}%
^{\prime},k_{y}-q_{_{y}}}\Big\{  [D_{n^{\prime}}^{\eta,s}D_{n}^{\eta
,s}]F_{nn^{\prime}}\left(  u\right)  +[C_{n^{\prime}}^{\eta,s}C_{n}^{\eta
,s}]F_{^{n-1,n^{\prime}-1}}\left(  u\right)  \Big\}   ^{2}, \label{A.11}%
\end{equation}
with $F_{nn^{\prime}}\left(  u\right) $ given after Eq. (7) in the text.

\end{document}